\pdfoutput=1
\documentclass[aps,prd,floatfix,twocolumn,reprint,superscriptaddress]{revtex4-1}
\usepackage{amsfonts}
\usepackage{mathrsfs}
\usepackage{amsmath}
\usepackage{color}
\usepackage{graphicx}
\usepackage{bm}
\usepackage{amssymb}
\usepackage{float}
\usepackage{xspace}
\usepackage{epstopdf}
\usepackage{dcolumn}
\usepackage{longtable}
\usepackage[colorlinks=true, letterpaper=true, pdfstartview=FitV, linkcolor=blue, citecolor=blue, urlcolor=blue]{hyperref}

\begin{document}
\title{Weyl Nodal Line-Surface Half-metal in CaFeO$_3$}

\author{Run-Wu Zhang}
\affiliation{Key Lab of advanced optoelectronic quantum architecture and measurement (MOE), Beijing Key Lab of Nanophotonics $\&$ Ultrafine Optoelectronic Systems, and School of Physics, Beijing Institute of Technology, Beijing 100081, China}

\author{Da-Shuai Ma}
\affiliation{Key Lab of advanced optoelectronic quantum architecture and measurement (MOE), Beijing Key Lab of Nanophotonics $\&$ Ultrafine Optoelectronic Systems, and School of Physics, Beijing Institute of Technology, Beijing 100081, China}

\author{Jian-Min Zhang}
\affiliation{Fujian Provincial Key Laboratory of Quantum Manipulation and New Energy Materials, College of Physics and Energy, Fujian Normal University, Fuzhou 350117, China}

\author{Yugui Yao}
\email{ygyao@bit.edu.cn}
\affiliation{Key Lab of advanced optoelectronic quantum architecture and measurement (MOE), Beijing Key Lab of Nanophotonics $\&$ Ultrafine Optoelectronic Systems, and School of Physics, Beijing Institute of Technology, Beijing 100081, China}
\date{\today}

\begin{abstract}
Manipulating the spin degrees of freedom of electrons affords an excellent platform for exploring novel quantum states in condensed-matter physics and material science. Based on ﬁrst-principles calculations and analysis of crystal symmetries, we propose a fully spin-polarized composite semimetal state, which is combined with the one-dimensional nodal lines and two-dimensional nodal surfaces, in the half-metal material CaFeO$_3$. In the nodal line-surface states, the Baguenaudier-like nodal lines feature six rings linked together, which are protected by the three independent symmetry operations: $\mathcal{PT}$, $\mathcal{M}_{y}$, and $\mathcal{\widetilde{M}}_{z}$. Near the Fermi level, the spin-polarized nodal surface states are guaranteed by the joint operation $\mathcal{T}\mathcal{S}_{2i}$ in the $k_{i(i=x,y,z)}=\pi$ plane. Furthermore, high-quality CaFeO$_3$ harbors ultra-clean energy dispersion, which is rather robust against strong triaxial compressional strain and correlation effect. The realization of the Weyl nodal line-surface half-metal presents great potential for spintronics applications with high speed and low power consumption.

\end{abstract}
\maketitle
\section{Introduction}
Spintronics is considered a rapidly developing field, using electron spin instead of its charge act as a go-between for data storage and transfer, has attracted extensive interest in recent research~\cite{1wolf2001spintronics,2vzutic2004spintronics}.
Among proposed new spintronic states~\cite{3de1983new,4van1995half,5ohno1996ga,6wang2008proposal,7kane2005z,8konig2007quantum,9hsieh2008topological,10wan2011topological,11chang2013experimental}, half-metals manifesting 100\% spin polarization are regarded as excellent candidates for promoting spin generation, injection, and transport~\cite{12park1998direct}. Particularly, it is a promising route that the electronic band structure of one spin channel exhibits symmetry protected band crossings near the Fermi level in a half-metal, then the semimetal states would be fully spin-polarized. The fully spin-polarized semimetals are characterized by the coexistence of topological and magnetic features. The search for desirable physical phenomena, such as fully spin-polarized fermion and topological order, could enable the realization of novel quantum devices and is therefore at the forefront of material science.

Akin to band structures of nonmagnetic semimetals, nodal states in half-metals fall into different categories via the dimensionality of the band crossing, including zero-dimensional (0D) nodal points, one-dimensional (1D) nodal loops and two-dimensional (2D) nodal surfaces, which makes the classification resemble nonmagnetic semimetals in three-dimensional (3D) systems~\cite{13wang2012dirac,14wang2013three,15huang2016topological,16chang2017type,17cao2017dirac,18weng2015weyl,19sun2015prediction,20soluyanov2015type,21ruan2016ideal,22autes2016robust,23ezawa2016loop,24li2017type,25wang2017antiferromagnetic,26zhang2018nodal,27ma2018mirror,28gong2018symmorphic,29yu2017nodal,30wang2017hourglass,31yan2017nodal,32sheng2017d,33zhang2018nodal,34zhong2016towards,35liang2016node,36wu2018nodal}. In 0D systems, Weyl nodal points in half-metals are characterized by the separated two-fold degenerate points in the Brillouin zone (BZ) and can lead to a Fermi arc on a certain surface. In contrast to 0D nodal points, 1D nodal loops with drumhead surface states in fully spin-polarized materials, provide more platforms for novel nodal states. Recent advances in the nodal loop half-metals offered a tremendous boost to the emerging field of nodal states in half-metals~\cite{37xu2011chern,38jiao2017first,39wang2018large,40zhang2018magnetization,41chen2019weyl,42wu2019weyl,43zhou2019fully,44wang2019two,45you2019two,46zhang2020nodal}. Such a state was recently predicted for Li$_3$(FeO$_3$)$_2$~\cite{41chen2019weyl}, which presents two fully spin-polarized nodal loops. Compared with Weyl nodal points and nodal loops in half-metal materials, the candidates of Weyl nodal surface half-metals~\cite{36wu2018nodal} are far fewer, and there is no associated experimental observation yet. An interesting question naturally arises: whether concrete materials can be realized to host both nodal surface and half-metal features? These materials are promising candidates for high-performance spin-based quantum devices in future.

Thus far, most materials with semimetallic characteristics are often suboptimal and suffer from various drawbacks. According to experimental studies and applications, the nontrivial band crossing located far away from the Fermi level and entangled with other irrelevant trivial bands, as well as the complex materials hard to implement experimentally, strongly limiting their applicability to high-efficiency nano-spintronics devices. The evolution of semimetal states from nonmagnetic to magnetic systems provides a brand-new platform to explore such intriguing physical properties. Therefore, seeking realistic “desired” nodal states in half-metals with fully spin-polarized nodal fermion will greatly refresh the designs of potential spintronic devices.

Overcoming aforementioned shortcomings, herein, based on ﬁrst-principles calculations and analysis of crystal symmetries, we propose and characterize a new Weyl nodal line-surface states in half-metal CaFeO$_3$. The CaFeO$_3$ shares a ferromagnetic ground state and generates 100\% spin polarization. The electronic states of spin-up channel manifest remarkable nodal states with the following characteristics: (i) the fully spin-polarized nodal lines resemble the “Baguenaudier” toy, which is composed of six nodal rings. These nodal lines are protected by the space-time inversion ($\mathcal{PT}$) and mirror ($\mathcal{M}_{y}$) as well as glide mirror ($\mathcal{\widetilde{M}}_{z}$) symmetries; (ii) the fully spin-polarized nodal surfaces appeared on the $k_{i(i=x,y,z)}=\pi$ plane near the Fermi level, dictated by the combination of nonsymmorphic twofold screw-rotational and time-reversal symmetries ($\mathcal{T}\mathcal{S}_{2i}$). The nodal line-surface states in half-metal mainly come from the orbitals of the light elements are rather robust against such tiny spin-orbit coupling (SOC). Moreover, the topological nodal line-surface states of CaFeO$_3$ can be effectively tailored by triaxial compressional strain and correlation effect. It is noteworthy that high-quality CaFeO$_3$ has been synthesized and possesses great application potential~\cite{47morimoto1997structure,48takeda2000metal,49woodward2000structural}. Thus, our results offer a realistic material platform for the exploration of the Weyl nodal line-surface half-metal states.

\section{Computational Method and Crystal Structure}

To systematically study the electronic properties of CaFeO$_3$, we perform the first-principles calculations by using the Vienna Ab initio Simulation Package (VASP) based on the generalized gradient approximation in the Perdew-Burke-Ernzerhof (PBE)~\cite{50perdew1996generalized,51blochl1994projector}. To obtain accurate crystalline structures of CaFeO$_3$, the advanced van der Waals density functional (\textit{i}.\textit{e}., SCAN+rVV10~\cite{52peng2016versatile}) is performed. In this work, the plane-wave cutoff energy is set as 560 eV, and $\Gamma$-centered \textit{k}-mesh with size $10\times7\times10$ is used for the primitive cell. To check the correlation effects of the transition metal iron atom, the DFT+\textit{U} method~\cite{53anisimov1991band,54dudarev1998electron} is carried out for calculating the band structures. Moreover, we construct the maximally localized Wannier functions (MLWFs) by employing the WANNIER90 code~\cite{55mostofi2008wannier90}. The corresponding surface states of nodal line half-metal are implemented with the WANNIERTOOLS package~\cite{56wu2018wanniertools} based on Green’s functions method.

\begin{figure}
\includegraphics[width=1.0\columnwidth]{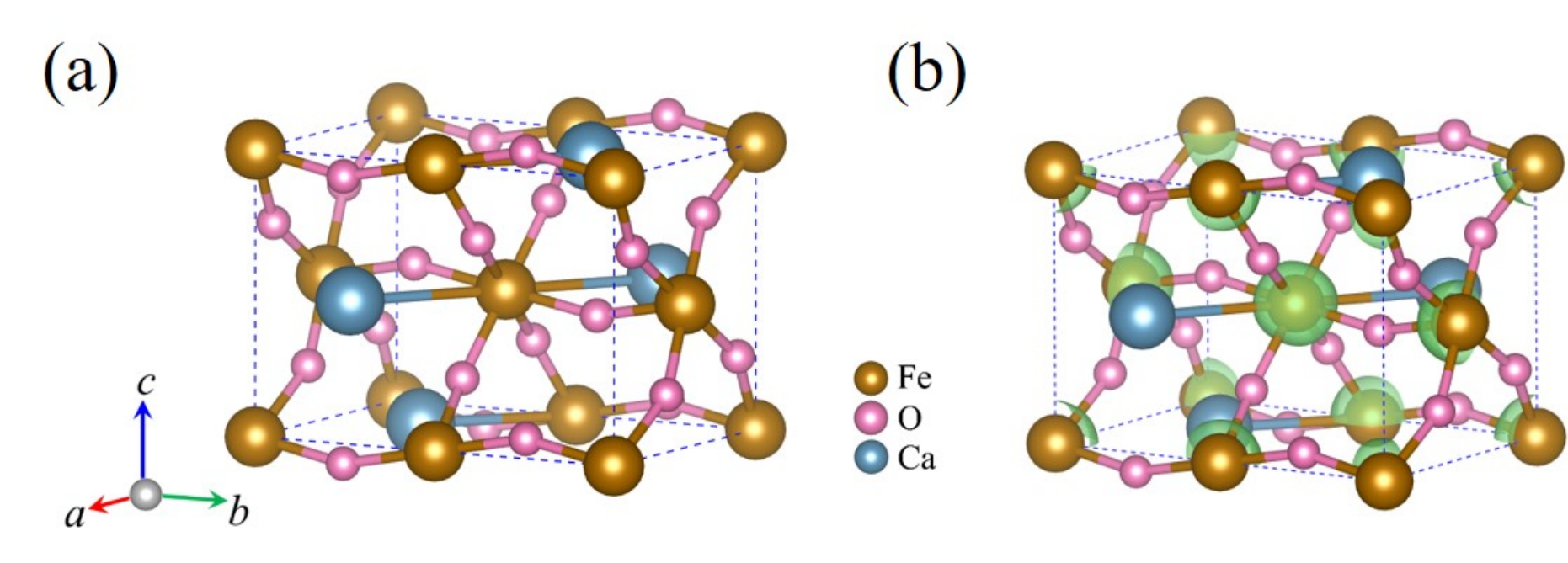}
\caption
{(a) The crystal structure and (b) 3D magnetic charge density of CaFeO$_3$. Note: gold, pink, and light blue color stand for iron, oxygen, calcium atoms, respectively.}
\label{fig1}
\end{figure}

The CaFeO$_3$ compound manifests the ABO$_3$ perovskite structure, usually crystallizes in ample phases under different experimental temperature circumstances such as the orthorhombic phase and monoclinic phase. At 300 K, simialr to the crystal structure of gadolinium orthoferrite (GdFeO$_3$)~\cite{49woodward2000structural}, CaFeO$_3$ is distorted from the ideal cubic perovskite crystal by the same octahedral tilting distortion, as illustrated in Fig.~\ref{fig1}(a). To optimize the orthorhombic phase CaFeO$_3$ more precisely, the advanced SCAN+rVV10 method is adopted. The calculated detailed structural parameters are summarized in Table \ref{Table1}. Hence, we take the optimized lattice structure by the SCAN+rVV10 method to further explore the electronic property. Remarkably, the optimized lattice parameters of CaFeO$_3$ are in excellent agreement with the experimental result~\cite{49woodward2000structural}. Furthermore, the Ca atoms occupy 4\textit{c} (0.5355, 0.7500, 0.9938) Wyckoﬀ position, Fe atoms occupy the 4\textit{a} (0.0, 0.0, 0.0) one, and O$_1$ atoms are at 8\textit{d} (0.7152, 0.0348, 0.2148) as well as O$_2$ atoms are at 4\textit{c} (0.5118, 0.2500, 0.5668). 
\begin{table}
  \caption{The lattice constants of the crystal CaFeO$_3$ (in unit of $\mathring{\mathrm{A}}$) are calculated based on the PBE and SCAN+rVV10 methods. The structural parameters of CaFeO$_3$ computed by the SCAN+rVV10 method are very close to the experimental results.}
  \label{Table1}
  \begin{tabular}{cccc}
    \hline
    \hline
    Phase	& Experiment~\cite{49woodward2000structural} & PBE   & SCAN+rVV10  \\
            & $a\setminus b\setminus c$ & $a\setminus b\setminus c$ & $a\setminus b\setminus c$\\
    \hline
    \textit{Pnma}	& $5.35\setminus 7.50\setminus 5.32$; & $5.43\setminus 7.59\setminus 5.36$; & $5.33\setminus 7.50\setminus 5.31$\\
    \hline
    \hline
  \end{tabular}
\end{table}

\section{Magnetic ground states and electronic structures}

In the orthorhombic phase CaFeO$_3$, the 3\textit{d} transition metal element Fe contributes the magnetic property, which is further verified by performing the 3D magnetic charge density,  as shown in Fig.~\ref{fig1}(b). Notably, all magnetic moments are localized around Fe atoms, which indicates that these Fe atoms can induce the magnetic property of CaFeO$_3$. Identifying the magnetic ground state of CaFeO$_3$ motivates us to explore the desirable properties. To check the magnetic ground states, we investigate four different kinds of possible magnetic orders, including ferromagnetic (FM) and three antiferromagnetic (AFM1, AFM2, and AFM3) configurations (see details in Appendix A). Consistent with earlier advances on the magnetic structure of CaFeO$_3$~\cite{57alexandrov2008ab,58cammarata2012spin,59dalpian2018bond}, our calculations indicate that the FM state possesses the lowest total energy among all possible magnetic states, meaning that the FM ground states are preferred. The magnetic moment is found to be 16 $\mu_{B}$ per primitive cell, which is mainly contributed by the four Fe atoms.

\begin{figure}
\includegraphics[width=0.8\columnwidth]{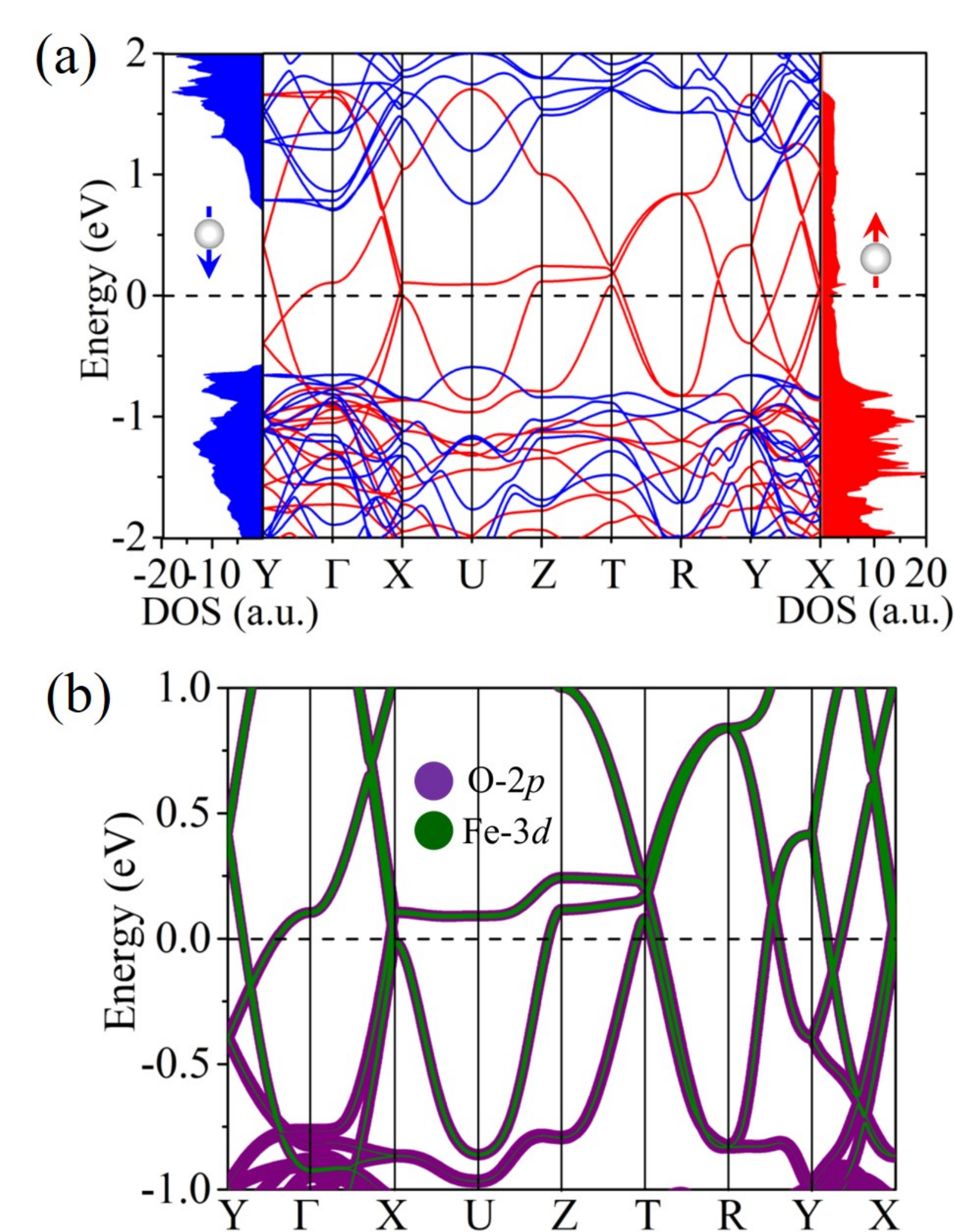}
\caption
{Spin-resolved band structure and the density of state of (a) CaFeO$_3$. Orbital-projected band structure of spin-up channel for (b) CaFeO$_3$ contributed by O-2\textit{p} and Fe-3\textit{d} (\textit{d}$_{xz}$ and \textit{d}$_{x^2-y^2}$) orbitals. Note: the red and blue arrows represent spin-up and spin-down channels, respectively. }
\label{fig2}
\end{figure}

From the projected density of states for FM order CaFeO$_3$, one notes that the CaFeO$_3$ behaves half-metallic state with spin-up channel possessing metallic while spin-down channel keeping semiconducting, as displayed in Fig.~\ref{fig2}(a). Considering the spin-polarized calculation, the band structure shows that the band crossing features around the Fermi energy are fully spin-polarized in the spin-up channel, while the spin-down channel exhibits a large gap of about 1.30 eV. The energy range of the spin-down channel is sufficient to overcome the interruption of irrelevant bands, therefore making ideal half-metals with fully spin-polarized. Moreover, the orbital-projection analysis shows that the obvious contribution to the conduction band minimum (CBM) and valence band maximum (VBM) near the Fermi level are mainly dominated by the Fe 3\textit{d} (\textit{d}$_{xz}$ and \textit{d}$_{x^2-y^2}$) and O 2\textit{p} orbitals, as plotted in Fig.~\ref{fig2}(b), which agrees well with previous studies~\cite{58cammarata2012spin}.

Buoyed by the attractive properties of topological nodal semimetals, we plot a new motif of topological nodal states in half-metal, termed as the Weyl nodal line-surface half-metal, which enjoys both the half-metal and semimetal traits, with fully spin-polarized Weyl fermions around Fermi level formed in a single spin channel. Via an enlarged view of the low-energy band structure in Fig.~\ref{fig2}(b), CaFeO$_3$ exhibits band-crossing and band-degeneracy: (i) the band crossings appear on the Y-$\Gamma$-X paths near the Fermi level; (ii) the bands are doubly degenerate along the high-symmetry X-U-Z-T paths. Before proceeding, it is notable that spin is a good quantum number in spin-polarized systems, and all Bloch states of spin and orbital can be reduced to two orthogonal subspaces in different spin channels~\cite{60chang2017topological,61wang2016time}. By choosing the axis along the spin polarization, the two spin channels are decoupled, which in turn preserves the crystal symmetries for the single spin channel. In the following, we further explore these charming phenomena.

\section{Weyl nodal line-surface states}

\begin{figure}
\includegraphics[width=1.0\columnwidth]{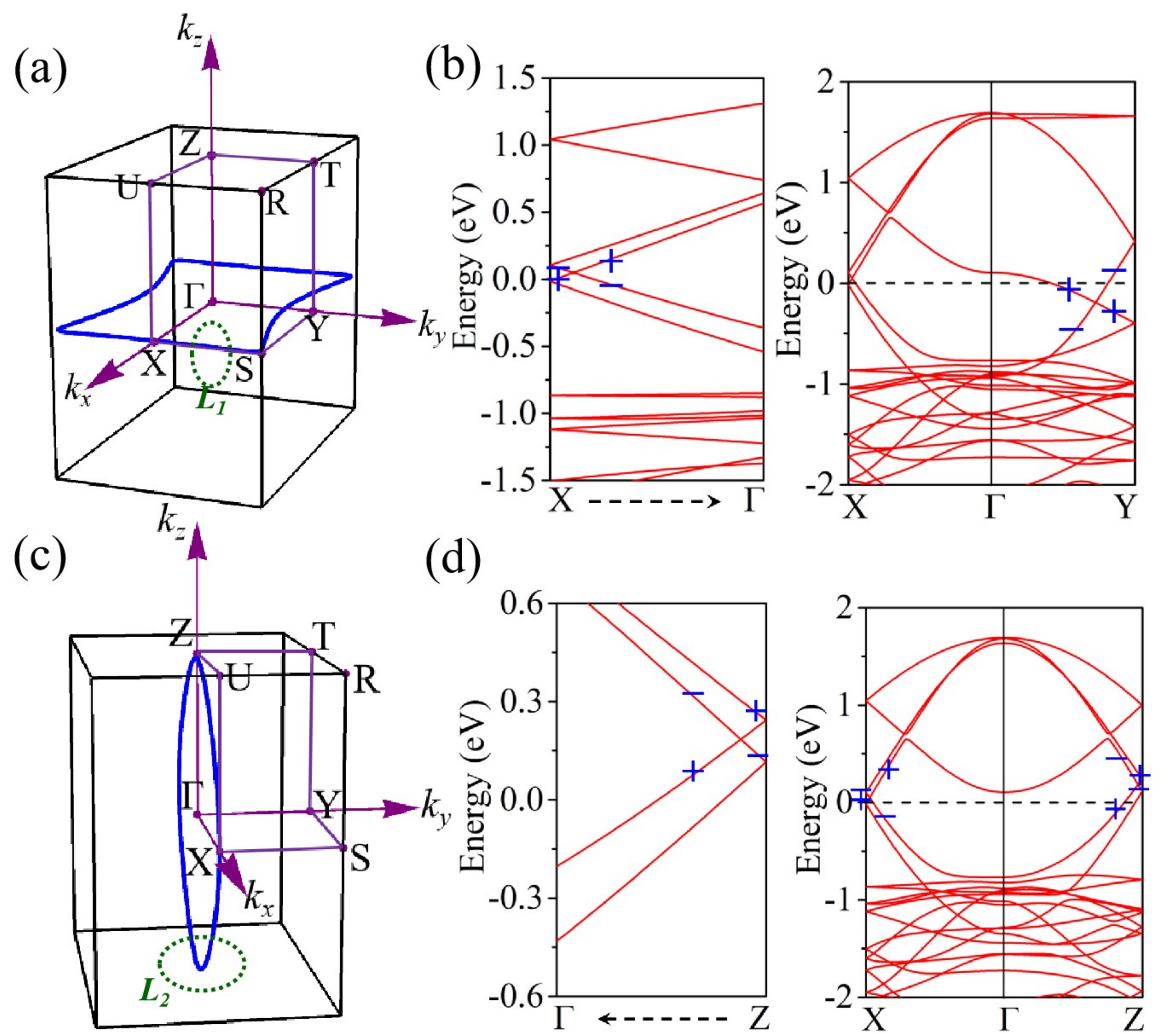}
\caption
{The 3D profile of the nodal lines lie on the (a) \textit{k}$_z$ = 0 plane and (c) \textit{k}$_y$ = 0 plane, respectively. Band structures of spin-up channel for CaFeO$_3$ along the X-$\Gamma$-Y path and X-$\Gamma$-Z path. The eigenvalues of
$\mathcal{\widetilde{M}}_{z}$ and $\mathcal{M}_{y}$ symmetry operators for each band is labeled $+$ and - in the enlarged band structures (b) and (d).}
\label{fig3}
\end{figure}

\begin{figure}
\includegraphics[width=1.0\columnwidth]{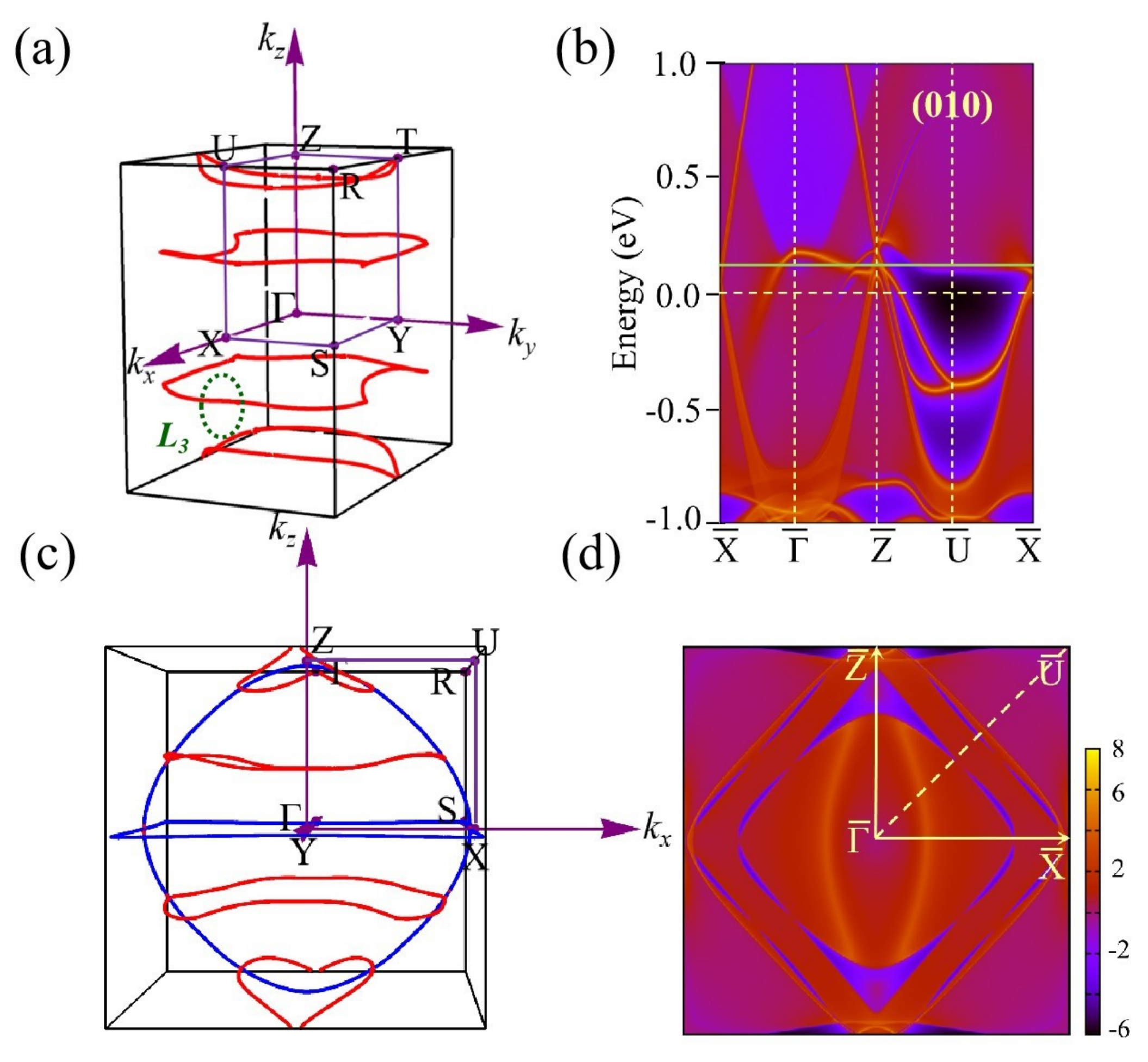}
\caption
{(a) The 3D view of the “snake”-like nodal lines in the Brillouin zone. The calculated Berry phase of an arbitrary green dashed ring (\textit{i}.\textit{e}., \textit{L}$_3$) intersecting a “snake”-like nodal line. (b) The projected surface states for the (010) surface. (c) The \textit{xoz} perspective of the nodal cage profiles in the Brillouin zone. (d) The isoenergy (\textit{E}$_f$ = 0.1 eV) surface around the $\overline{\Gamma}$ point for the (010) surface.}
\label{fig4}
\end{figure}

Regarding the space group \textit{Pnma} in CaFeO$_3$, it is important to note the presence of the following symmetry operation: the inversion $\mathcal{P}$, mirror reflection $\mathcal{M}_{y}:(x,y,z) \rightarrow (x,-y+\frac{1}{2},z)$, two glide mirror reflections $\mathcal{\widetilde{M}}_{x}:(x,y,z) \rightarrow (-x+\frac{1}{2},y+\frac{1}{2},z+\frac{1}{2})$ and $\mathcal{\widetilde{M}}_{z}:(x,y,z) \rightarrow (x+\frac{1}{2},y,-z+\frac{1}{2})$ as well as three twofold screw rotations $\mathcal{S}_{2x}:(x,y,z) \rightarrow (x+\frac{1}{2},-y+\frac{1}{2},-z+\frac{1}{2})$, $\mathcal{S}_{2y}:(x,y,z) \rightarrow (-x,y+\frac{1}{2},-z)$ and $\mathcal{S}_{2z}:(x,y,z) \rightarrow (-x+\frac{1}{2},-y,z+\frac{1}{2})$. We first consider the linear band-crossings features of CaFeO$_3$ along the X-$\Gamma$-Y path on the \textit{k}$_z$ = 0 plane. As depicted in Fig.~\ref{fig3}(a), we perform a scan of the momentum distribution for these band-crossings and observe a continuous nodal line lie exactly on the \textit{k}$_z$ = 0 plane, demonstrating these crossing points are not isolated. In Fig.~\ref{fig3}(b), the CBM and VBM have opposite signs of the eigenvalues of $\mathcal{\widetilde{M}}_{z}$ concering the \textit{k}$_z$ = 0 plane, indicating the gapless nodal line is protected by the $\mathcal{\widetilde{M}}_{z}$ symmetry. A similar analysis also can be applied to the \textit{k}$_y$ = 0 plane, as plotted in Fig.~\ref{fig3}(c) and Fig.~\ref{fig3}(d). Therefore, two mutually perpendicular nodal lines in CaFeO$_3$ are exactly lying on the two mirror invariant planes (\textit{k}$_y$ = 0 and \textit{k}$_z$ = 0). Moreover, through careful scanning calculation, we reveal that the trace of the band-crossings actually forms “snake”-like nodal lines (not necessary to lie on the mirror invariant planes), as illustrated in Fig.~\ref{fig4}(a). Analogous to alkaline-earth compounds~\cite{15huang2016topological}, such “snake”-like nodal lines are protected by $\mathcal{PT}$ symmetry. To further validate the topological properties of CaFeO$_3$, we introduce the Berry phase calculation along with distinctive lines. We find that, Berry phase equals to $\pi$ along a ring \textit{L}$_1$ (\textit{L}$_2$ and \textit{L}$_3$) threading the nodal lines located on the $\mathcal{M}_{y}$, $\mathcal{\widetilde{M}}_{z}$ as well as “snake”-like lines. Noticeably, due to these topological nodal lines derived from a single spin channel, these low-energy nodal-line fermions behave fully spin-polarized features, which will be useful for spintronics applications.

Unique fully spin-polarized nodal line states of CaFeO$_3$ inspire us to further explore potential applications. To directly present the intriguing properties of topological nodal lines, the hallmark drumhead surface states are performed. Fig.~\ref{fig4}(b) shows the surface spectrum for the (010) surface. As for Fig.~\ref{fig4}(c), we plot the nodal line profiles in the \textit{xoz} perspective. Similar to the longitude and latitude of the earth, a large loop linked other small loops and each has a clear track. The isoenergy band contour of the surface at the (010) surface is obtained in Fig.~\ref{fig4}(d), which is basically consistent with the case of Fig.~\ref{fig4}(c) (the constant energy slice at \textit{E}$_f$ = 0.1 eV). More importantly, the 2D nearly ﬂat drumhead-like surface states in nodal line half-metals not only have a large density of states but also are fully spin-polarized, which provide a good playground for spintronics~\cite{1wolf2001spintronics,2vzutic2004spintronics}, topological superconductivity~\cite{62kopnin2011high,63volovik2015standard}, and topological catalysts~\cite{64rajamathi2017weyl,65li2018topological}.

\begin{figure}
\includegraphics[width=1.0\columnwidth]{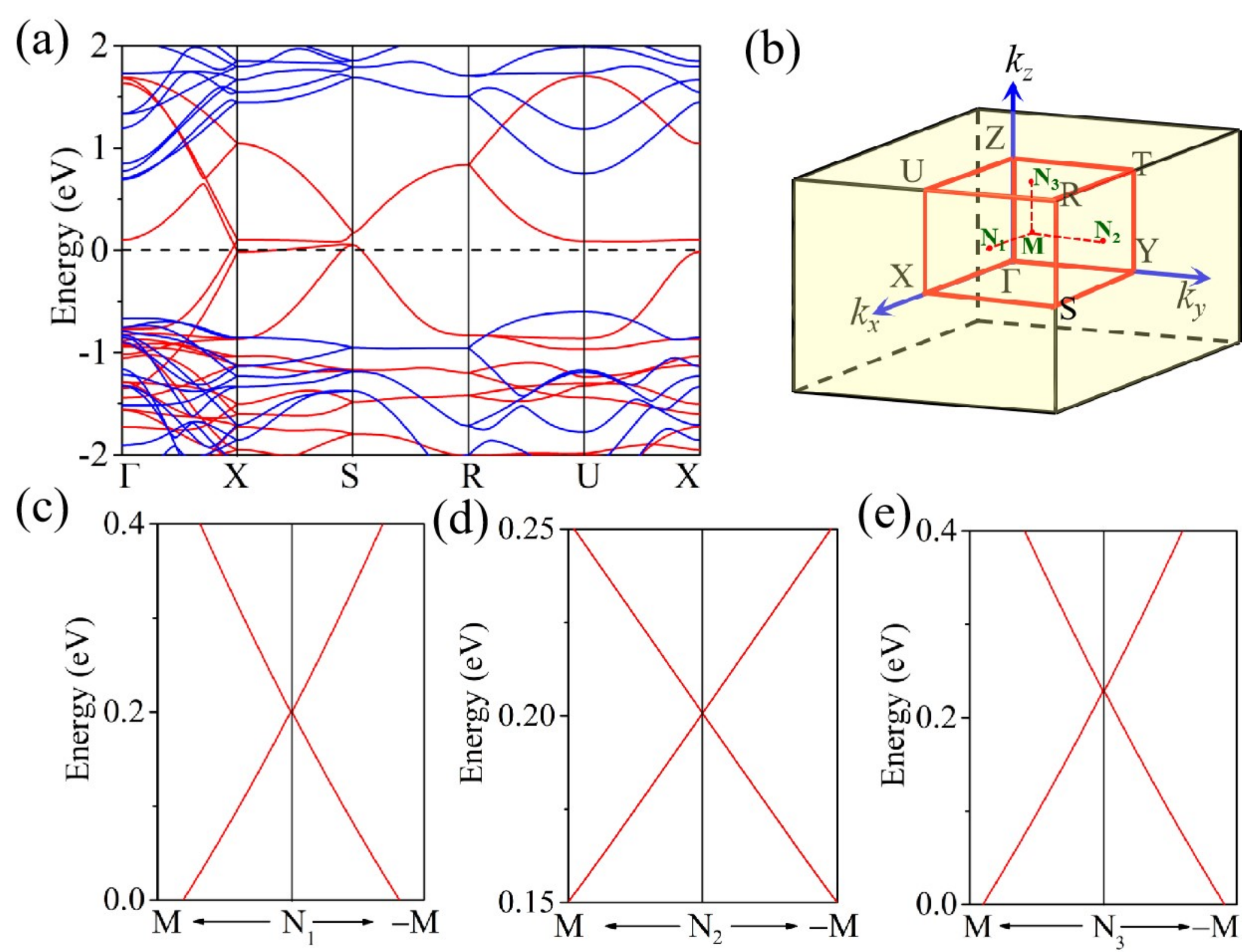}
\caption
{(a) Spin-resolved band structure of CaFeO$_3$. (b) The corresponding Brillouin zone. Note: M is the middle point of $\Gamma$-R path and N$_1$ (N$_2$ and N$_3$) is the middle point of X-R (Y-R and Z-R) path. Enlarged band dispersion along the M and N$_1$ (c) (N$_2$ (d) and N$_3$ (e)) path, showing the linear crossing between two bands along the \textit{k}$_x$ (\textit{k}$_y$ and \textit{k}$_z$) direction.}
\label{fig5}
\end{figure}

Next, we switch to fully spin-polarized nodal surface states of CaFeO$_3$. In the absence of SOC, the spin and orbital degrees of freedom are decoupled,
such that each spin channel can be considered in the subspace. Considering the spin polarization, we first discuss the degeneracy along the X-S-R-U-X path, as seen in Fig.~\ref{fig5}(a). The twofold screw rotation $\mathcal{S}_{2x}:(x,y,z) \rightarrow (x+\frac{1}{2},-y+\frac{1}{2},-z+\frac{1}{2})$, which is a nonsymmorphic symmetry involving a half translation along the rotation axis. In momentum space, $\mathcal{S}_{2x}$ inverses \textit{k}$_y$ and \textit{k}$_z$ while preserves \textit{k}$_x$. One finds that $(\mathcal{S}_{2x})^2$ = \textit{T}$_{100}$ = $e^{-ik_{x}}$, where \textit{T}$_{100}$ is the translation along the \textit{x}-direction by a lattice constant. $\mathcal{T}$ is antiunitary and inverses \textit{k} with $\mathcal{T}^2$ = 1. Thus, $\mathcal{T}\mathcal{S}_{2x}$ is antiunitary and only inverses \textit{k}$_x$. Since $\left[\mathcal{T}, S_{2x}\right]=0$, $\mathcal{T}\mathcal{S}_{2x}$ satisfies $(\mathcal{T}\mathcal{S}_{2x})^2$ = $e^{-ik_{x}}$. When \textit{k}$_x$ = $\pi$, each \textit{k} point is invariant under the $\mathcal{T}\mathcal{S}_{2x}$ operation, and $(\mathcal{T}\mathcal{S}_{2x})^2$ = -1 means that there is a Kramers-like degeneracy due to the antiunitary $\mathcal{T}\mathcal{S}_{2x}$ symmetry. Thus, the bands in the \textit{k}$_x$ = $\pi$ plane must be doubly degenerate, forming the nodal surface. A similar argument has been proposed in the previous discussion of the material CsCrI$_3$~\cite{36wu2018nodal}. Moreover, in Fig.~\ref{fig5}(c), we check the dispersion along a generic \textit{k}-path M-N$_1$, which is not a high-symmetry path (see Fig.~\ref{fig5}(b)). The linear crossing can be obtained from the generic path, which further proved the existence of nodal surface states. It is notable that the symmetry requires the presence of the nodal surface and confines the location of the nodal surface in the \textit{k}$_x$ = $\pi$ plane. It puts no constraint on the energy and the dispersion of the nodal surface. For the nodal surface state in \textit{k}$_x$ = $\pi$ plane, the varition of the energy disperse in a narrow range near the Fermi level. Similar to the case of \textit{k}$_x$ = $\pi$ plane, the nodal surface states can be revealed in the \textit{k}$_y$ = $\pi$ and \textit{k}$_z$ = $\pi$ planes~\cite{66yu2019circumventing}. The protection mechanisms of the other two nodal surfaces are consistent with the first one, and similar features are plotted in Fig.~\ref{fig5}(d) and Fig.~\ref{fig5}(e).

\section{Discussion and conclusion}
 
Considering the correlation effects of the transition metal \textit{d} orbitals, we perform the GGA+\textit{U} calculations by taking into account the on-site Coulomb interaction, and find that the nodal line and nodal surface states are stable within \textit{U} values (from 3 to 7 eV). It is also stable against a wide range of strains. The details are presented in Appendixes. Thus, such nodal line-surface states are robust against correlation and strain effects.

In conclusion, we identify that CaFeO$_3$ hosts the fully spin-polarized nodal line-surface states. The ferromagnetic ground state of CaFeO$_3$ is half-metal and features nodal line-surface states in the spin-up channel. The fully spin-polarized nodal line states possess “Baguenaudier”-like rings due to the $\mathcal{PT}$, $\mathcal{M}_{y}$, and $\mathcal{\widetilde{M}}_{z}$ symmetries with negligible SOC. Moreover, the drumhead-like fully spin-polarized surface states are obtained. Also, CaFeO$_3$ holds unique nodal surface states with fully spin-polarized as protected by joint operation $\mathcal{T}\mathcal{S}_{2i}$ at the $k_{i(i=x,y,z)}=\pi$ plane. To our knowledge, such coexistence of spin-polarized nodal cage and nodal surface states is different from the current proposed nodal line or surface in nonmagnetic systems, and this unique candidate may have great potential in future spintronics applications. 

\section{Acknowledgements}

This work is supported by the National Natural Science Foundation of China (Grants Nos. 12047512, 11734003, 12061131002), the National Key R\&D Program of China (Grant No. 2020YFA0308800, 2016YFA0300600), the Strategic Priority Research Program of Chinese Academy of Sciences (Grant No. XDB30000000). R.W.Z. also acknowledge the support from the Project Funded by China Postdoctoral Science Foundation (Grant No.2020M680011).

R.W.Z and D.S.M. contributed equally to this work.




\bibliography{ref}

\end{document}